\documentstyle[sprocl]{article}

\def\sst{\scriptscriptstyle}

\def\beq{\begin{equation}}
\def\eeq{\end{equation}}
\def\beqa{\begin{eqnarray}}
\def\eeqa{\end{eqnarray}}
\newcommand{\da}{\dot{a}}
\newcommand{\db}{\dot{b}}

\newcommand{\dda}{\ddot{a}}

\newcommand{\pa}{a^{\prime}}
\newcommand{\pb}{b^{\prime}}

\newcommand{\ppa}{a^{\prime \prime}}

\newcommand{\fda}{\frac{\da}{a}}

\newcommand{\fdb}{\frac{\db}{b}}

\newcommand{\fdda}{\frac{\dda}{a}}

\newcommand{\fpa}{\frac{\pa}{a}}
\newcommand{\fpb}{\frac{\pb}{b}}

\newcommand{\fppa}{\frac{\ppa}{a}}

\newcommand{\nc}{\newcommand}
{\nc{\lsim}{\mbox{\raisebox{-.6ex}{~$\stackrel{<}{\sim}$~}}}
{\nc{\gsim}{\mbox{\raisebox{-.6ex}{~$\stackrel{>}{\sim}$~}}}

\begin{document}

\title{COSMOLOGICAL EXPANSION IN THE RANDALL-SUNDRUM WARPED
COMPACTIFICATION}

\author{J.M.\ CLINE}

\address{McGill University, 3600 University St.\\
Montr\'eal, Qu\'ebec H3A 2T8, Canada\\
E-mail: jcline@physics.mcgill.ca} 

\maketitle\abstracts{The cosmology of a brane-universe embedded in 
a higher dimensional bulk spacetime presents some peculiarities
not seen in ordinary ($3+1$) dimensional gravity.  I summarize the
current understanding, with emphasis on the suggestion by Randall and
Sundrum that the bulk is 5-D anti-deSitter space, leading to a solution
of the weak scale hierarchy problem.} 

\section{Large Versus Small Extra Dimensions}

In the last few years there has been a revival of interest in the idea of
extra dimensions, first proposed by Kaluza and Klein.  The new
realization of Arkani-Hamed, Dvali and Dimopoulos (ADD) was that the extra
dimensions could be macroscopically large if one assumed that our ($3+1$)
dimensional universe is a slice (a 3-brane) of the higher dimensional
bulk.\cite{ADD}\ The particles of the standard model should be restricted
to the brane so that no light Kaluza-Klein (KK) excitations exist, which
otherwise would have already been seen.  Gravity, however, can propagate
in the extra bulk dimensions (otherwise they would have no observable
consequences whatsoever).  The effect of the extra dimensions can only be
seen on distance scales less than the order of their size.  With $N$ extra
compact dimensions of size $R$, Newton's gravitational force law for
two masses $m_1$ and $m_2$, separated by a distance $r$, is modified to
\beq
\label{eq:NNewton}
F = {\Gamma(\frac{3+N}{2})\over 4\pi^{(3+N)/2}}\left(
{m_1 m_2 \over M^{2+N} r^N}\right),\qquad r\ll R,
\eeq
where $M$ is the new quantum gravity scale appearing in the
Einstein-Hilbert action for gravity in $4+N$ dimensions,
\beq
\label{eq:action}
	S = - \frac12 M^{2+N}\int d^4\!x\, d^{N}\!y \sqrt{|g|} {\cal R},
\eeq
and $y_{\sst I}$ parametrize the extra dimensions.  At larger separations, the
gravitational flux is no longer diluted by spreading out in the bulk, so
the force reverts to its usual form,
\beq
\label{eq:Newton}
F = \frac{1}{8\pi}\left({m_1 m_2 \over M_p^{2} r^2}\right),\qquad r\gg R,
\eeq
involving the 4-D Planck mass $M_p$.  Deviations of (\ref{eq:Newton}) from
its $1/r^2$ form have only been tested at separations greater than a
millimeter or so, showing that $R$ could be as large as 1 mm.  This is
obviously far bigger than the limit which exists if the standard
model particles are allowed to propagate in the bulk, 
$R \lsim 10^{-3}$ fm.

The relationship between the $4$-D and the $(4+N)$-D gravity scales can
easily be deduced by requiring that the action (\ref{eq:action}) reduce to
the usual one after integrating over the extra dimensions.  Let us suppose
that the line element in the full spacetime has the simple form
\beq
\label{eq:metric}
ds^2 = a^2(y_{\sst I})\left( dt^2 - \sum_{i=1}^3 dx_i dx_i\right) - b^2(y_{\sst I}) 
	\sum_{I=1}^N dy_{\sst I} dy_{\sst I}.
\eeq
Then, if the brane is located at $y_{\sst I}=0$, the relationship is
\beq
\label{eq:Mrel}
	M_p^2 = M^{N+2} \int 
\left({a(y_{\sst I})\over a(0)}\right)^2 b^N\!(y_{\sst I})\, d^N\! y.
\eeq

In ADD, the geometry is assumed to be factorizable, meaning that $a$
does not depend upon $y_{\sst I}$; hence the integral just gives the volume of
the extra dimensions: $M_p^2 = M^{N+2}V_N$.  Since $V_N$ can be quite
large (mm$^3$), this has the interesting consequence that $M$ could be
at the TeV scale, yet be consistent with the much higher scale of
$M_p$.  This opens the mind-boggling possibility that all new particle
physics, including quantum gravity, could become accessible at the
LHC.  Moreover we have a partial explanation of the weak scale
hierarchy problem, the question of why $M_p$ is 16 orders of magnitude
larger than the $W$ boson mass.  It is not really a solution because
one is left with the annoying question of why $R$ is so much larger
than the natural scale, $1/M$.  The exact size depends on the number of
extra dimensions.  If $N=1$ it is not possible to obtain $M\sim 1$ TeV
because $R$ is too large; demanding that $R = 1$ mm gives $M\sim 10^8$
GeV.  But for $N=2$, the TeV scale emerges just as the experimental
bound on $R$ is saturated, and for higher dimensions it can be attained
with smaller sizes, $R\sim 100$ fm in the case of $N=6$.

The experimental constraints on large extra dimensions come from the
effects of the KK excitations of the graviton, which can be very light,
$m_n = n/R$ for integer $n$.  The only thing which saves these
particles from being easily discovered is their weak interactions; like
the ordinary graviton, their couplings are suppressed by $1/M_p^2$ (as
opposed to $1/M^2$).  Consequently the bounds from accelerator physics
are rather weak: $M\gsim$ several TeV.\cite{ADDbounds}\ Astrophysics
gives better constraints, at least for $N=1$ and 2.  One such bound
comes from requiring that supernova 1987A not cool too quickly by
graviton emission,\cite{SNbounds}\ giving $M \gsim 100$ TeV for $N=1$
and $M\gsim 5$ TeV for $N=2$.  In the early universe, KK gravitons can
be produced by thermal processes, and decay slowly into photons that
would distort the cosmic gamma ray background unless $M$ obeys bounds
similar to the supernova ones.\cite{Hall}\

Actually, the last-mentioned bound is quite generous toward the ADD
scenario because it assumes that, by some miracle, the universe is
already free from primordially produced KK gravitons at temperatures
near 1 MeV--a necessary condition since otherwise the gamma rays
produced by their decays would destroy deuterium and consequently the
successful predictions of big bang nucleosynthesis.  It is quite
difficult to justify this assumption.  Benakli and Davidson showed that
if $M=1$ TeV, the reheat temperature after inflation would have to be
no greater than 0.1 GeV, for $N=6$; for smaller $N$ the bound is even
more stringent.\cite{BD}\ This is well below what is needed for
electroweak baryogenesis, which is generally considered to be the
lowest temperature mechanism available.  Therefore baryogenesis
presents a major challenge to the ADD idea.

Randall and Sundrum (RS) have suggested another way of solving the
hierarchy problem with an extra dimension,\cite{RS}\ which avoids the
difficulties encountered by ADD.  They considered just a single extra
dimension, compactified on an orbifold $S_1/Z_2$, a circle modded by
$Z_2$.  The coordinate is in the range $y\in [-1,1]$, with the
endpoints identified and with $y\leftrightarrow -y$ being the orbifold
symmetry.  One places a 3-brane at each of the orbifold fixed points,
$y=0$ and $y=1$.  They have equal and opposite tensions, $\pm\sigma$
(tension is the 4-D energy density, which has the same form as a 4-D
cosmological constant).  In addition there is a 5-D cosmological constant
in the bulk, $\Lambda$.  The stress-energy tensor is therefore
\beq
\label{eq:stress}
	T_{\mu\nu} = (g_{\mu\nu} - n_\mu n_\nu) \sigma \left(\delta(y) - 
	\delta(y-1)\right)/b + \Lambda g_{\mu\nu},
\eeq
where $n_\mu$ is the normal to the branes (hence the brane tensions make
no contribution to $T_{yy}$).  A static solution to the 5-D Einstein
equations exists if 
\beq
\label{eq:RScond}
	\Lambda = - {\sigma^2\over 6 M^3},
\eeq
and it has the form of eq.\ (\ref{eq:metric}) with 
\beq
\label{eq:RSsoln}
	a(y) = e^{-kby}; \qquad k = |\Lambda/\sigma|
\eeq
and $b$, the size of the extra dimension, being undetermined.  Using eq.\
(\ref{eq:Mrel}), one finds that $M_p$ is related to the 5-D gravity scale
by 
\beq
	M_p^2 = {M^3\over k}(1-e^{-2kb}).
\eeq

The dramatic consequence of this solution is that if one considers the
Lagrangian for a particle confined to the brane at $y=1$, it takes its
canonical form only after a Weyl rescaling of the field by the ``warp
factor'' $a(1) = e^{-kb}$.  If the Lagrangian originally had $M_p$ as the
mass scale for the particle, it becomes rescaled by $\exp(-kb)$.  One can
take all the parameters $M$, $\Lambda$, $\sigma$ and $k$ to be of order
$M_p$ to the appropriate power; then if $b\sim 36/k \sim 36/M_p$, one
obtains TeV scale masses on the $y=1$ brane (henceforth called the TeV
brane).  Clearly $bk \sim 36$ is a much more moderate hierarchy than
$M_p/M \sim 10^{16}$, so this constitutes an attractive possible
explanation of the weak scale.  Furthermore the extra dimension is still
small, so the KK gravitons can be sufficiently heavy to present no  
difficulties in the early universe. 

This solution to the hierarchy problem requires that we are living on the
negative tension brane (taken to be at $y=1$).  The positive tension brane
at $y=0$ has no such suppression of its masses, so it is referred to
as the Planck brane, and must constitute a kind of hidden sector.

\section{Effect of Extra Dimensions on Cosmological Expansion}

For a while it appeared that cosmology could provide an interesting
constraint on large extra dimensions.  Bin\'etruy, Deffayet and Langlois
(BDL) considered the cosmological expansion of 3-brane universes in a 5-D
bulk and found solutions in which the Hubble expansion rate in the brane
was related to the energy density $\rho$ on the brane by \cite{BDL}
\beq
\label{eq:wrongrate}
	H = \fda= {\rho\over 6 M^3},
\eeq
in contrast to the usual Friedmann equation, $H\propto \sqrt{\rho}$. 
Although other authors had found inflationary solutions with
this property,\cite{others}\ BDL were the first to point out that it would
be a problem for later cosmology.  Especially, such a modification to the
expansion rate would probably drastically alter the predictions of big
bang nucleosynthesis.

It is not difficult to see from the 5-D Einstein equations, $G_{\mu\nu}
= M^{-3} T_{\mu\nu}$,  why one gets
the unusual dependence of $H$ on $\rho$.  Consider the $00$ component, 
\beq
\label{eq:ein00}
\fda \left( \fda+ \fdb \right) = \frac{a^2}{b^2}
\left(\fppa + \fpa \left( \fpa - \fpb \right) \right) + {1\over 3 M^3}
T_{00}. 
\eeq
To obtain the delta functions in $T_{00}$, eq.\ (\ref{eq:stress}), $a'(y)$
must be discontinuous at both branes, and the discontinuity is
proportional to the total energy density on the branes. Moreover the
orbifold symmetry (as well as common sense) requires that $a(y)$ be
symmetric about either brane, so that $a'(1-\epsilon) = -a'(1+\epsilon)$. 
This implies that $a'(y)$ itself is linearly proportional to the brane
tension $\sigma$.   Now consider the $yy$ component, which has no delta
functions:
\beq
\label{eq:ein55}
	\left(\fda\right)^2 + \fdda =  2\frac{a^2}{b^2}
	 \left(\fpa\right)^2 - {1\over 3 M^3}T_{yy}
\eeq
Recalling that $H = \fda$, clearly $H^2$ will get contributions
proportional to $(a'/a)^2\sim \sigma^2$ as well as $\Lambda$.  In fact, if
we allow for a cosmological constant in the bulk and 
extra energy densities $\rho_{\!\sst P}$ and $\rho_{\sst T}$, in addition to
the respective tensions $\sigma$ and $-\sigma$ on the Planck and TeV
branes, the complete expression becomes\footnote{the factor $e^{-2kb}$
was first pointed out by ref.\ \cite{CGRT}}\ 
\beq
\label{eq:newrate}
	H^2 = {(\sigma+\rho_{\!\sst P})^2\over 36 M^6} + {\Lambda\over 6M^3}
	    = {(-\sigma+e^{-2kb}\rho_{\sst T})^2\over 36 M^6} + {\Lambda\over 
	6M^3}
\eeq
It was noticed\,\cite{CGKT,CGS1} that by tuning $\sigma$ to cancel the
contribution from $\Lambda$ in the limit $\rho_i = 0$, one obtains an
expression for $H$ which at leading order in $\rho$ has the desired
$\sqrt{\rho}$ form, plus small fractional corrections of order $\rho/M^4$. 
Not surprisingly, in retrospect, this tuning is precisely the same
condition (\ref{eq:RScond}) required by RS to obtain their solution.  (Any
deviation from this condition results in an effective 4-D cosmological
constant and therefore a nonstatic solution.)  But at the time we first
noticed this coincidence, it was striking to us, since we were unaware of
RS and had thus come upon the condition (\ref{eq:RScond}) starting from a
completely different motivation from that of RS. 

However, all is not well with the cosmological solution leading to eq.\
(\ref{eq:newrate}).  For one thing, the energy densities on the two
branes are constrained, $\rho_{\sst T} = -e^{2kb}\rho_{\!\sst P}$.
Moreover, this implies that $\rho_{\sst T} < 0$, {\it i.e.,} that the
energy density of matter on the TeV brane is negative, a physically
unacceptable situation.  Thus, although cosmology appeared to be normal
on the Planck brane (for densities $\rho_{\!\sst P} \ll M^4$), not so
on the TeV brane, where the hierarchy problem is solved.  This seemed
to present a problem for the RS proposal.\cite{CGKT,CGS1}

There were several attempts to solve this problem.  In one it was observed
that by decompactifying the orbifold,\cite{RS2} 
placing the TeV brane at the position required by the
hierarchy problem ($y\cong 36/kb$), and giving it a tension between $0$
and $-\sigma/2$, one could obtain the normal Hubble rate on the TeV brane
with a positive energy density.\cite{CGS2}\ However this solution involved
simultaneous expansion of the bulk, which is unacceptable for late time
cosmology because a growing $b(t)$ leads to a Planck mass which is
increasing in time, according to eq.\ (\ref{eq:Mrel}).  Hence gravity
would be getting weaker on the TeV brane, contrary to stringent
constraints on the time variation of Newton's constant.  In another
attempt it was pointed out that the normal Friedmann equation would ensue
if the ${yy}$ component of $T_{\mu\nu}$ was allowed to have a rather
complicated dependence on the bulk coordinate $y$, rather than being a
constant ($\Lambda$).\cite{Olive1}\ The origin of such a
dependence seemed obscure (but see below).

It was recently shown that both of the cosmological problems of brane
universe models---the artificial relation between the energy densities on
the two branes, and the generically ``wrong'' form for the Friedmann
equation---can be solved by introducing a mechanism for insuring that the
size of the extra dimension remains fixed while the branes
expand.\cite{CGRT}\ Recall that this degree of freedom was completely
undetermined in the RS model, meaning that it corresponds to a modulus,
{\it i.e.,} a massless field, in this context called the radion.  This is
problematic in itself, because it implies a fifth force, as in
scalar-tensor theories of gravity, which in the present case has couplings
suppressed by the TeV rather than the Planck scale.\cite{GW3}\ In the
absence of a mechanism for stabilizing this modulus (see ref.\ \cite{GW2}\
for such a mechanism), it is not surprising that a fine-tuning between the
brane energies as in eq.\ (\ref{eq:newrate}) should be needed to insure
that the bulk does not expand along with the branes.\cite{Ell} 

Somewhat less obvious is the fact that the normal Friedmann equation also
results if $b$ (the size of the compact dimension) is stabilized.
Heuristically, this occurs because the bulk cosmological constant
$\Lambda$ is now replaced by a potential for $b$, $V(b)$.  Since the 
$yy$ component of Einstein's equation comes from the variation of the
action with respect to $b$, a new term appears in $T_{yy}$,
\beq
	T_{yy} = b^2( V(b) + bV'(b)).
\eeq
Therefore the $G_{yy}$ equation, which we used in the argument above to
obtain $H\propto\rho$, is no longer available for fixing the magnitude
of $H$; rather it determines $b$,\cite{JC} because $b$ no longer sits
at the bottom of the potential during a period of cosmological
expansion, but is slightly shifted.  Moreover ref.\ \cite{Olive2}
showed that the $y$-dependent stress-energy needed in \cite{Olive1} to
get the correct Friedmann equation automatically arises from the
stabilization of the radion.

In \cite{CGRT} it is shown that, if the radion is stabilized, then as long
as the excess energy densities $\rho_i$ (above and beyond those needed in
eq.\ (\ref{eq:RScond}) to get a static solution) are small compared to the
cutoff scales ($M_p$ on the Planck brane and 1 TeV on the TeV brane), 
the two branes expand at approximately the same rate, given by
\beq
	H^2 = {8\pi G\over 3}\left(\rho_{\!\sst P} + e^{-4kb}\rho_{\sst T} +
	\int_0^1 \! dy\, b\, e^{-4kby} \rho_{\rm bulk}(y)\right).
\eeq
This expression can be derived from the effective 4-D Lagrangian obtained
by integrating over the extra dimension in the background of the RS
metric.  Notice the factor of $e^{-4kb}$ multiplying $\rho_{\sst T}$.  This is
precisely the same redshifting of mass scales that applies to all masses
on the TeV brane.  Thus $\rho_{\sst T}$ represents the bare value of the energy
density, presumably of order $M_p^4$, while $e^{-4kb}\rho_{\sst T}$ is the
physically observed value.  No such suppression occurs for matter living
on the Planck brane.  Therefore, if the expansion of the universe is to be
dominated by matter on our (TeV) brane, it is necessary to demand that the
Planck brane (and the bulk) be devoid of matter.  Fortunately this does
not seem to pose a major challenge: inflation will effectively empty out
the Planck brane, as long as it harbors no nearly massless particles and
the reheat temperature is significantly below the cutoff.  Although it is
tempting to suggest that $\rho_{\!\sst P}$ could be the dark matter of the
universe, it is difficult to see how it could be made sufficiently small,
if it is not zero. 

\section{Outlook}

The Randall-Sundrum proposal for solving the hierarchy problem with
a small extra dimension looks compatible with most cosmological
requirements.  Unlike the ADD scenario of large extra dimensions, it
does not suffer from the problem of light KK gravitinos wreaking havoc
with nucleosynthesis and the cosmic gamma ray background.  Furthermore
it might have a plausible string theoretic origin,\cite{CGS3}\
perhaps being related to the 5D-anti-deSitter space/conformal field theory
correspondence and the holographic principle.\cite{Verlinde}\  There
remain a few puzzles.  One is the apparent possibility of a
``dark radiation'' term in the Friedmann equation,\cite{darkrad}
\beq
	H^2 = {8\pi G\over 3}\left(\hbox{usual terms} + {{\cal C}\over 
	a^4}\right)
\eeq
which arises as an initial condition, due to the fact that the
solutions to 5-D gravity have an additional constant of integration
relative to 4-D.  Does this oddity also disappear when the extra dimension
is stabilized?  Another problem is inflation, which typically requires the
presence of an intermediate scale, $M_i \sim 10^{13}$ GeV, to get the
right magnitude of density perturbations, $\delta\rho/\rho \sim M_i/M_p$.  
No such intermediate scale exists if the TeV scale is the true cutoff on
our brane.  A third interesting question is how to generalize the
RS scenario to higher dimensions, which is just beginning to be 
explored.\cite{CGS3,higherD}

\end{document}